\documentclass[12pt]{article}
\usepackage{graphicx}
\begin{document}
\centerline{\bf Majority-vote on directed Barab\'asi-Albert networks}

\bigskip
\centerline{F.W.S. Lima}

\bigskip
\noindent
Departamento de F\'{\i}sica,
Universidade Federal do Piau\'{\i}, 57072-970 Teresina - PI, Brazil

\medskip
  e-mail: wel@ufpi.br
\bigskip

{\small Abstract: On directed Barab\'asi-Albert networks with two and
 seven  neighbours selected by each added site, the Ising model was seen
 not to show a spontaneous magnetisation. Instead, the decay time for
 flipping of the magnetisation followed an Arrhenius law for Metropolis
 and Glauber algorithms, but for Wolff cluster flipping the
 magnetisation  decayed exponentially with time. On these networks the
 Majority-vote model with noise is now studied through Monte Carlo
simulations.
 However, in this model,  the order-disorder phase transition of the
 order parameter
 is well defined in this system. We calculate the value of the critical
 noise parameter $q_{c}$ for several values of connectivity $z$ of the
 directed Barab\'asi-Albert network. The critical
 exponentes $\beta/\nu$, $\gamma/\nu$ and
 $1/\nu$ were calculated for
 several values of $z$.}


\bigskip

 {\bf Introduction}

 It has been argued that nonequilibrium stochastic spin systems on
 regular square lattice with up-down symmetry fall in the universality
 class of the equilibrium Ising model \cite{g}. This conjecture was
 found in several models that do not obey detailed balance \cite{C,J,M}.
 Campos $et$ $al$. \cite{campos} investigated the majority-vote model
 on small-world network by rewiring the two dimensional square lattice. These
 small-world networks, aside from presenting quenched disorder, also
 posses long-range interactions. They found that the critical exponents
 $\gamma/\nu$ and $ \beta/\nu$ are different from the Ising model and depend
 on the rewiring probability. However, it was not evident that the
 exponent change was due to the disordered nature of the network or due to
 the presence of long-range interactions. Lima $et$ $al$. \cite{lima0}
 studied the majority-vote model on Voronoi-Delaunay random lattices
 with periodic boundary conditions. These lattices posses natural quenched
 disorder in their conecctions. They showed that presence of quenched
 connectivity disorder is enough to alter the exponents $\beta/\nu$
 and $\gamma/\nu$ from the pure model and therefore that is a relevant
term to
 such non-equilibrium phase-transition.
 Sumour and Shabat \cite{sumour,sumourss} investigated Ising models on
 directed Barab\'asi-Albert networks \cite{ba} with the usual Glauber
 dynamics.  No spontaneous magnetisation was
 found, in contrast to the case of undirected  Barab\'asi-Albert networks
 \cite{alex,indekeu,bianconi} where a spontaneous magnetisation was
 found lower a critical temperature which increases logarithmically with
 system size. More recently, Lima and Stauffer \cite{lima} simulated
 directed square, cubic and hypercubic lattices in two to five dimensions
 with heat bath dynamics in order to separate the network effects from
 the effects of directedness. They also compared different spin flip
 algorithms, including cluster flips \cite{wang}, for
 Ising-Barab\'asi-Albert networks. They found a freezing-in of the
 magnetisation similar to  \cite{sumour,sumourss}, following an Arrhenius
 law at least in low dimensions. This lack of a spontaneous magnetisation
 (in the usual sense)
 is consistent with the fact
 that if on a directed lattice a spin $S_j$ influences spin $S_i$, then
 spin $S_i$ in turn does not influence $S_j$,
% following remark added
 and there may be no well-defined total energy. Thus, they show that for
 the same  scale-free networks, different algorithms give different
 results. Now we study the Majority-vote model on directed
 Barab\'asi-Albert network and different from the Ising model, the
 order-disorder phase transition of
 order parameter well it is defined in this system. We calculate the
 $\beta/\nu$, $\gamma/\nu$, and $1/\nu$ exponents and  these are
 different from the Ising model and depend on the values of
 connectivity $z$ of the
 directed Barab\'asi-Albert network.

\bigskip

\begin{figure}[hbt]
\begin{center}
\includegraphics[angle=-90,scale=0.46]{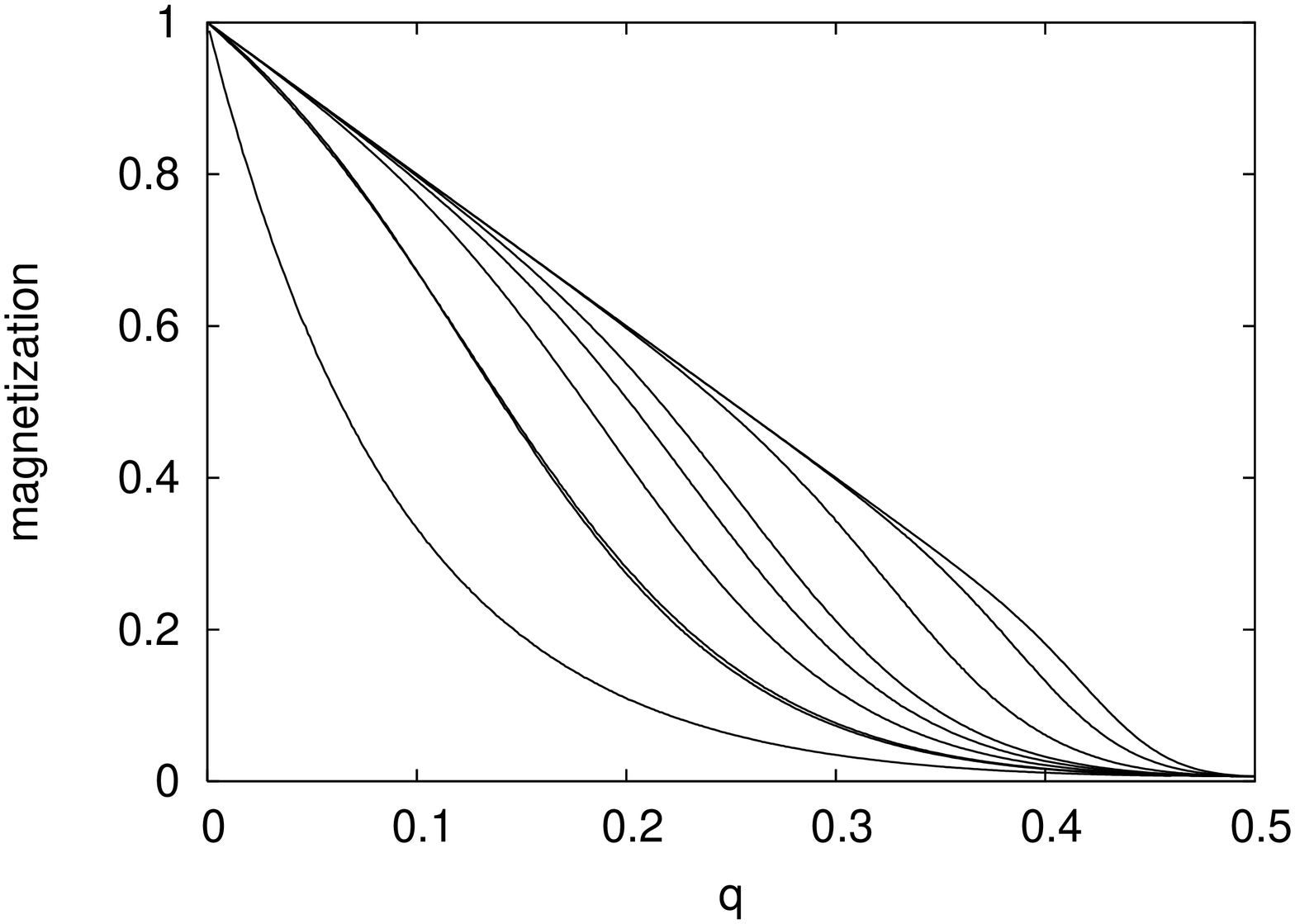}
\includegraphics[angle=-90,scale=0.46]{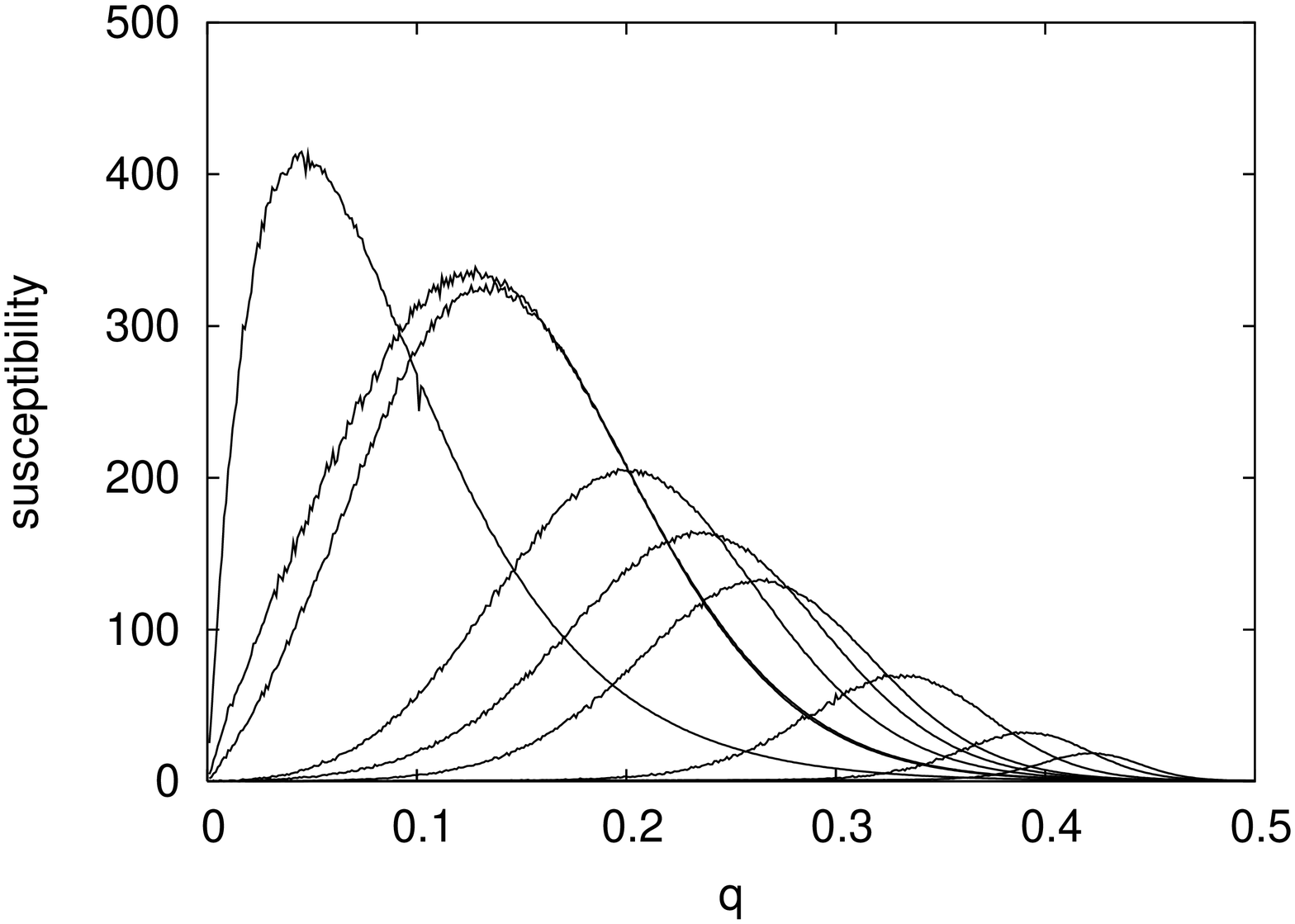}
\end{center}
\caption{
Magnetisation and susceptibility as a function of the noise parameter $q$,
for
$N=16000$ sites. From left to ri,ht, $z=2$, $3$, $4$, $6$, $8$, $10$,
$20$, $50$,
and $100$ .}
\end{figure}

\bigskip

\begin{figure}[hbt]
\begin{center}
\includegraphics[angle=-90,scale=0.46]{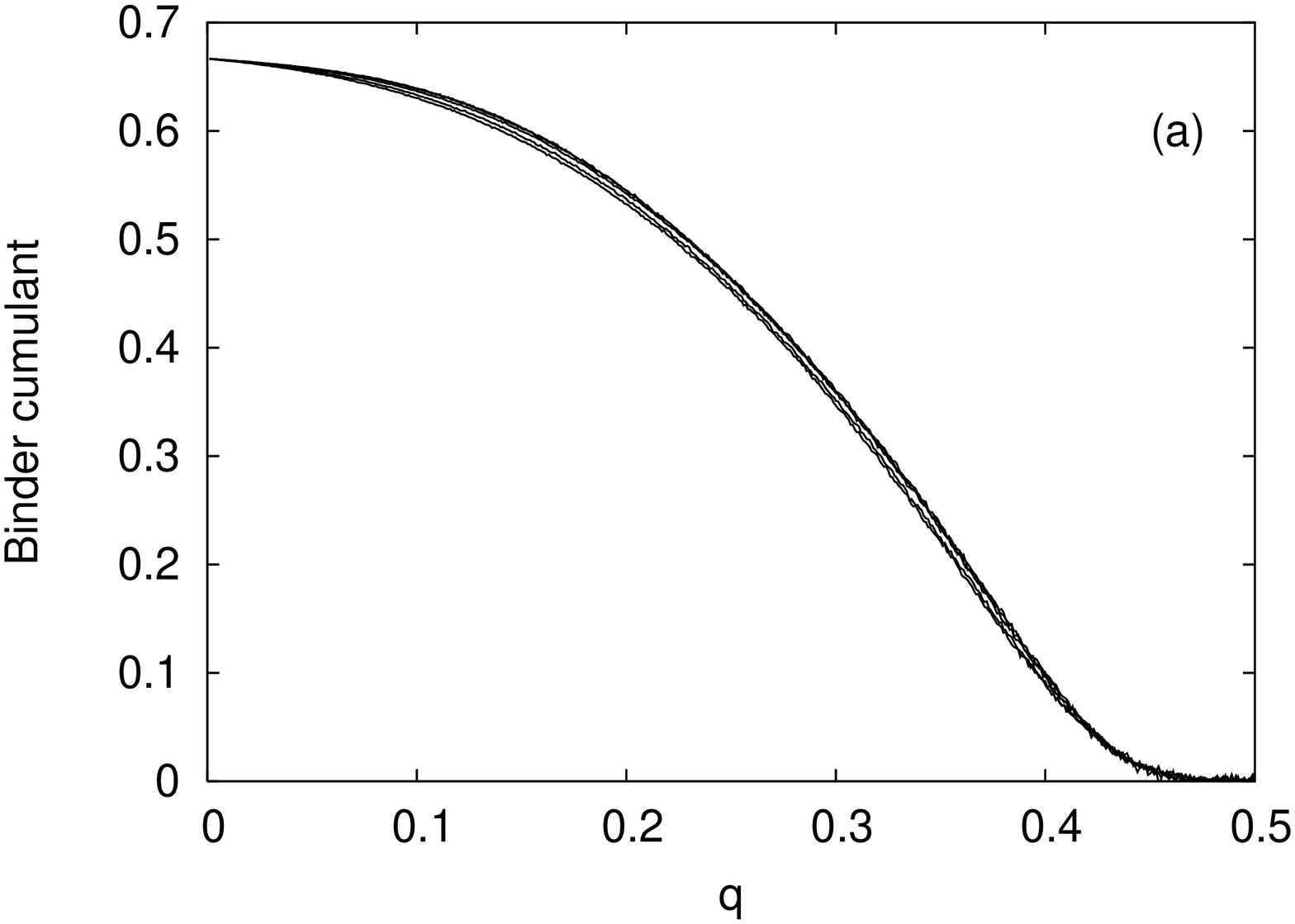}
\includegraphics[angle=-90,scale=0.46]{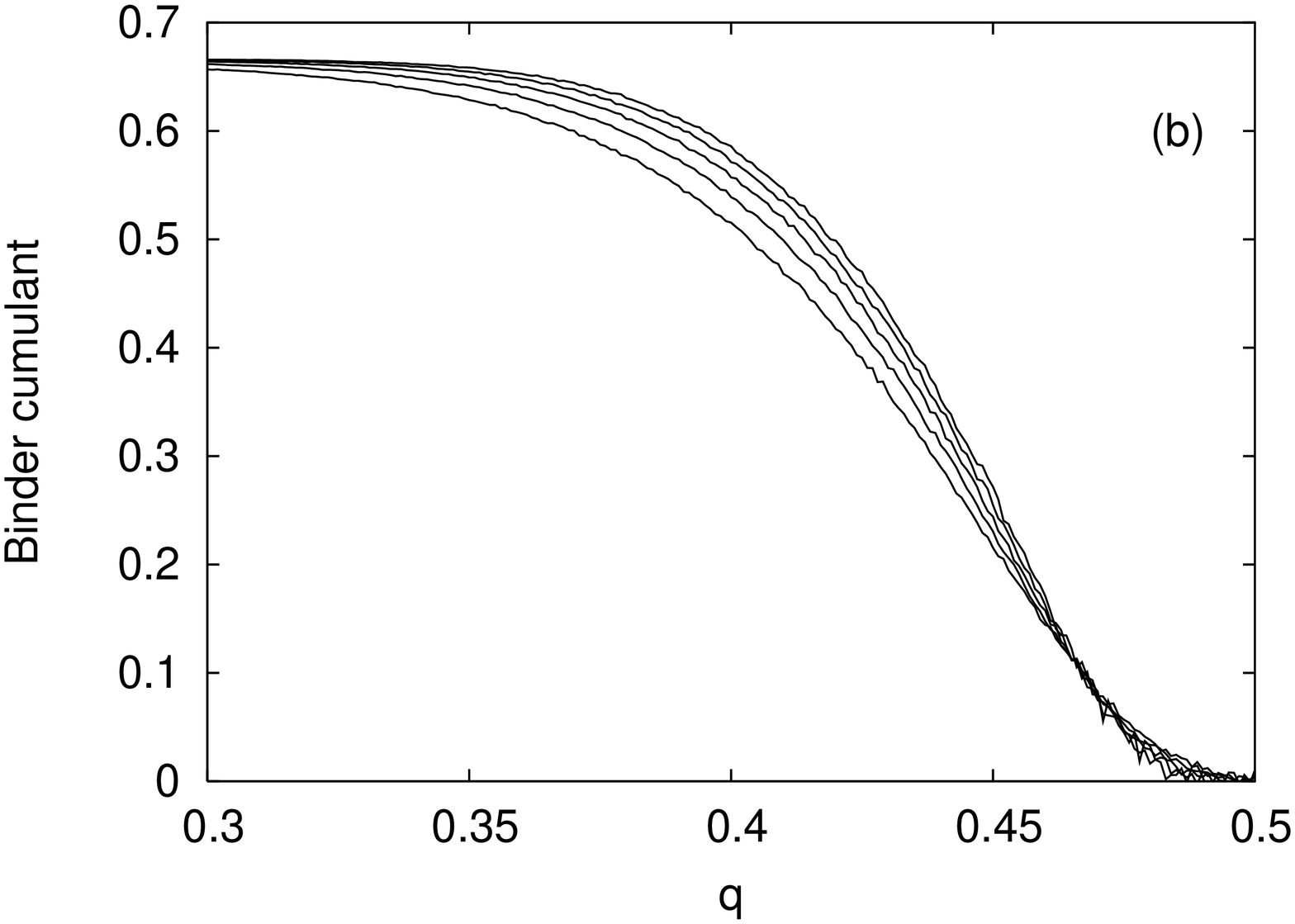}
\end{center}
\caption{
Binder's fourt-order cumulant as a function of $q$. In part (a) we have
$z=3$ and
part (b) $z=50$.}
\end{figure}

\begin{figure}[hbt]
\begin{center}
\includegraphics[angle=-90,scale=0.60]{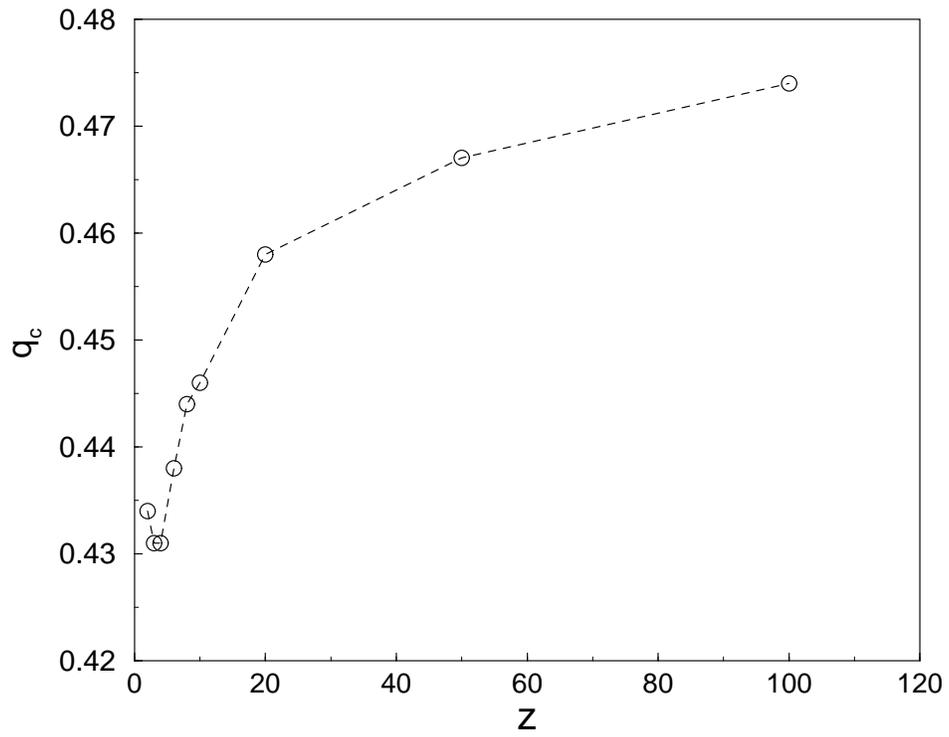}
\end{center}
\caption{The phase diagram, showing the dependence of critical
noise parameter $q_{c}$ on connectivity $z$.
}
\end{figure}

\begin{figure}[hbt]
\begin{center}
\includegraphics[angle=-90,scale=0.60]{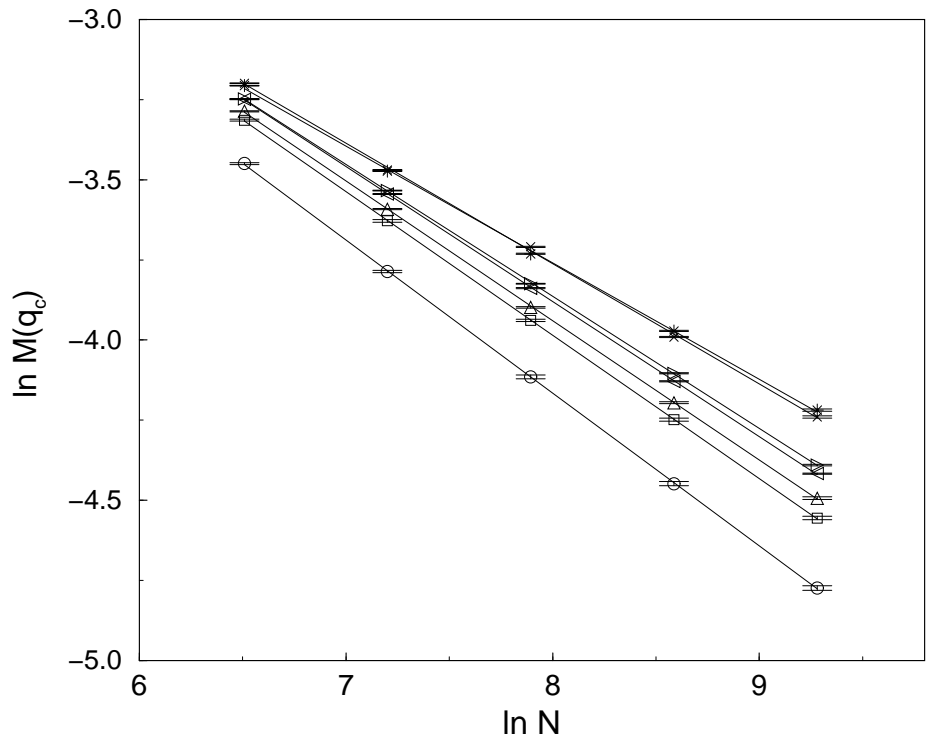}
\end{center}
\caption{$\ln M(q_{c})$ versus $\ln N$. From bottom to top, $z=2$, $4$, $6$,
$10$, $20$, $50$, and $100$. }
\end{figure}

\bigskip

\begin{figure}[hbt]
\begin{center}
\includegraphics[angle=-90,scale=0.60]{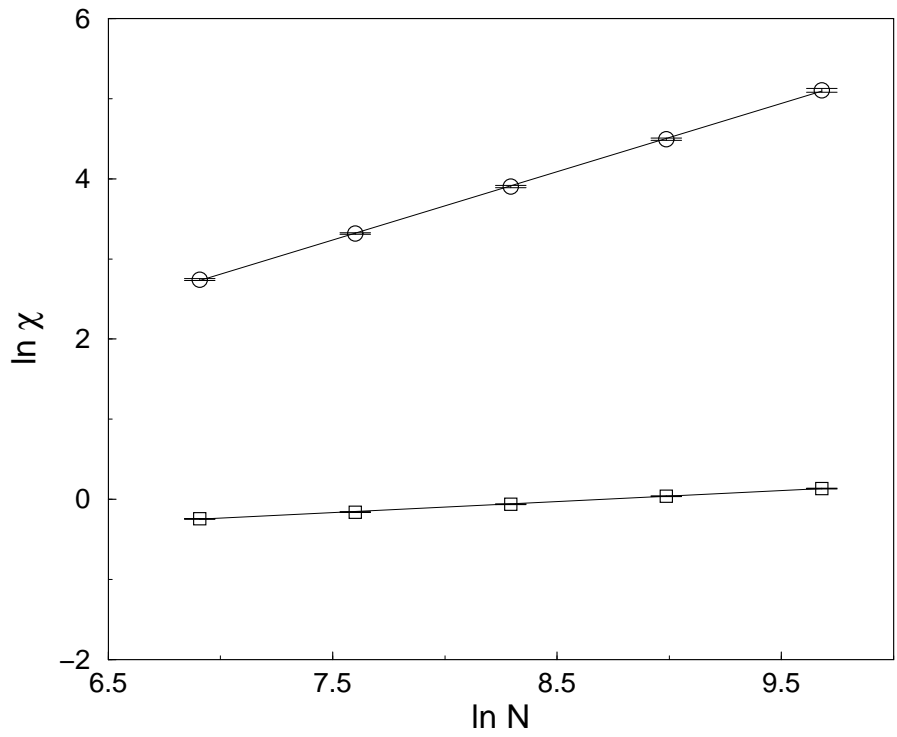}
\end{center}
\caption{Plot of ln $\chi^{max}(N)$ (circle) and ln$ \chi(q_{c})$ (square)
versus ln
$N$ for
connectivity $z=8$.
}
\end{figure}
\bigskip

\begin{figure}[hbt]
\begin{center}
\includegraphics[angle=-90,scale=0.60]{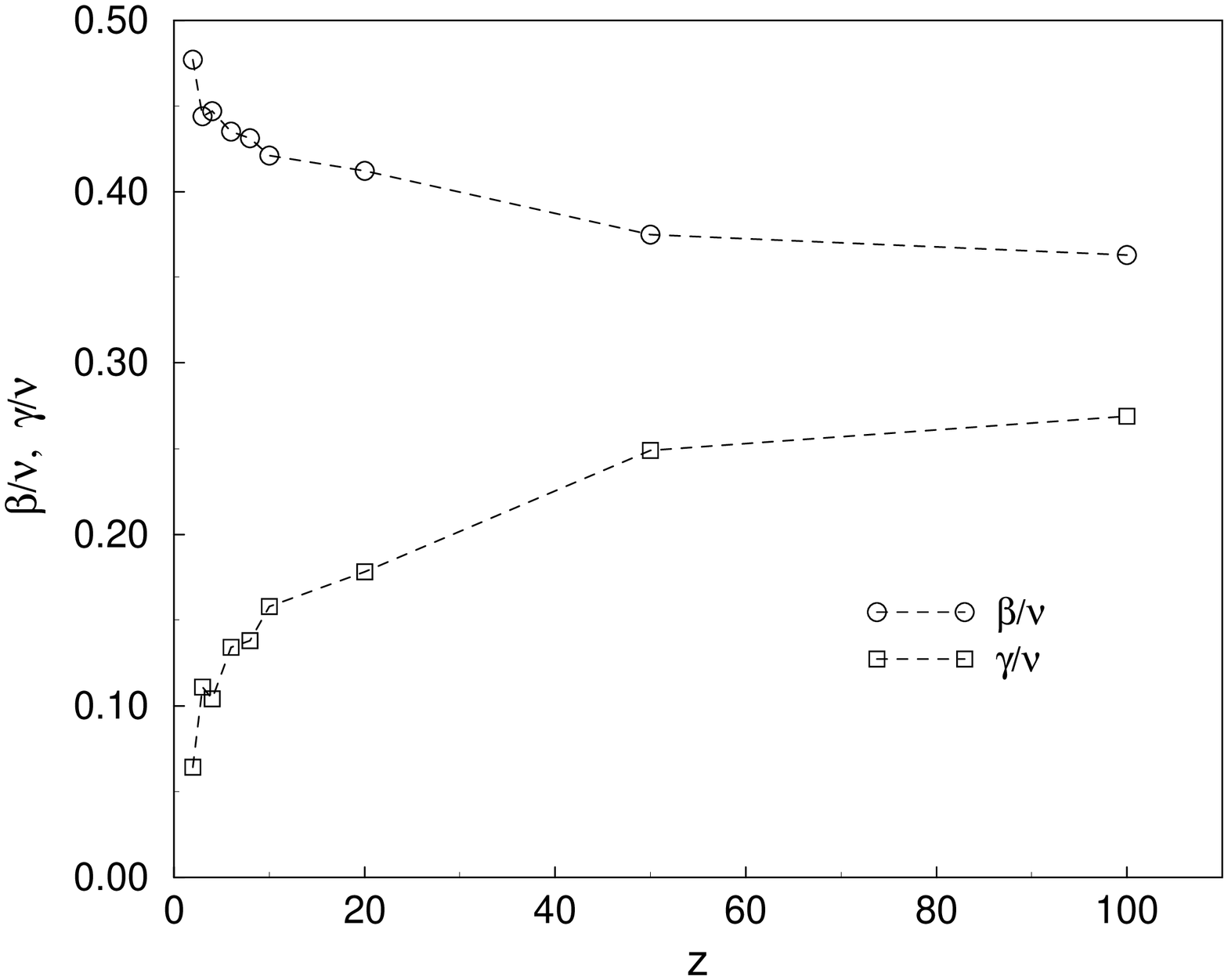}
\end{center}
\caption{ Critical behavior the $\beta/\nu$ and $\gamma/\nu$ exponents as
a function of
connectivity $z$.}
\end{figure}

\bigskip

{\bf Model and Simulaton}

We consider the majority-vote model, on directed
Barab\'asi-Albert Networks, defined \cite{mario,jff,lima0,brad} by a set of
"voters" or spins variables ${\sigma}$ taking the values $+1$ or
$-1$, situated on every site of a directed
Barab\'asi-Albert Networks with $N$ sites, and evolving in time by single
spin-flip
like dynamics with a probability $w_{i}$ given by
\begin{equation}
%\begin{center}
w_{i}(\sigma)=\frac{1}{2}\biggl[ 1-(1-2q)\sigma_{i}S\biggl(\sum_{\delta
=1}^{k_{i}}\sigma_{i+\delta}\biggl)\biggl],
%\end{center}
\end{equation}
where $S(x)$ is the sign $\pm 1$ of $x$ if $x\neq0$, $S(x)=0$ if $x=0$,
and the
sum runs over all nearest neighbors of $\sigma_{i}$. In this network, each
new site
added to the network selects $z$
already existing sites as neighbours influencing it; the newly
added spin does not influence these neighbours. The control parameter
$q$ plays the role of the temperature in equilibrium systems and measures
the probability of aligning antiparalle to the majority of neighbors.

To study the critical behavior of the model we define the variable
$m=\sum_{i=1}^{N}\sigma_{i}/N$. In particular , we were interested in the
magnetisation, susceptibility and the reduced fourth-order cumulant:
\begin{equation}
M(q)=[<|m|>]_{av},
\end{equation}
\begin{equation}
\chi(q)=N[<m^2>-<|m|>^{2}]_{av},
\end{equation}
\begin{equation}
U(q)=\biggl[1-\frac{<m^{4}>}{3<m^2>^{2}}\biggl]_{av},
\end{equation}
where $<...>$ stands for a thermodynamics average and $[...]_{av}$ square
brackets
for a averages over the 20 realizations.

These quantities are functions of the noise parameter $q$ and obey the
finite-size
scaling relations

\begin{equation}
M=N^{-\beta/\nu}f_{m}(x)[1+ ...],
\end{equation}
\begin{equation}
\chi=N^{\gamma/\nu}f_{\chi}(x)[1+...],
\end{equation}
\begin{equation}
\frac{dU}{dq}=N^{1/\nu}f_{U}(x)[1+...],
\end{equation}
 where $\nu$, $\beta$, and $\gamma$ are the usual critical
exponents, $f_{i}(x)$ are the finite size scaling functions with
\begin{equation}
x=(q-q_{c})N^{1/\nu}
\end{equation}
being the scaling variable, and the brackets $[1+...]$ indicate
corretions-to-scaling terms. Therefore, from the size dependence of $M$
and $\chi$
we obtained the exponents $\beta/\nu$ and $\gamma/\nu$, respectively.
The maximum value of susceptibility also scales as $N^{\gamma/\nu}$.
Moreover, the
value of $q$ for which $\chi$ has a maximum, $ q_{c}^{\chi_{max}}=q_{c}(N)$,
is expected to scale with the system size as
\begin{equation}
q_{c}(N)=q_{c}+bN^{-1/\nu},
\end{equation}
were the constant $b$ is close to unity. Therefore, the  relations $(7)$
and $(9)$
are used to determine the exponente $1/\nu$. We have checked also if the
calculated
exponents satisfy the hyperscaling hypothesis
\begin{equation}
2\beta/\nu+\gamma/\nu=D_{eff}
\end{equation}
in order to get the effective dimensionality, $D_{eff}$, for various
values of $z$.

We have performed Monte Carlo simulation on directed Barab\'asi-Albert
networks with
various values of connectivity $z$. For a given $z$, we used systems
of size $N=1000$, $2000$, $4000$, $8000$, and $16000$. We waited $10000$
Monte Carlo
steps (MCS) to make the system reach the steady state, and the time
averages were
estimated from the next $ 10000$ MCS. In our simulations, one MCS is
accomplished
after all the $N$ spins are updated. For all sets of parameters, we have
generated
$20$ distinct networks, and have simulated $20$
independent runs for each distinct network.

\bigskip

{\bf Results and Discussion}

In Fig. 1 we show the dependence of the magnetisation $M$  and the
susceptiblity
$\chi$  on the noise parameter, obtained from simulations on directed
Barab\'asi-Albert network with $16000$ sites and several values of
conectivity $z$.
In the part (a) each  curve for $M$, for a given value of  $N$ and $z$,
suggests
that there is  a phase transition from an ordered state to a disordered
state. The
phase transition occurs at a value of the critical noise parameter
$q_{c}$, which is
an increasing function the conectivity $z$  of the directed Barab\'asi-Albert
network. In the part (b) we
show the corresponding behavior of the susceptibility $\chi$, the value 
of $q$
where  $\chi$ has a maximum is here identified as $q_{c}$. In Fig. 2  we
plot the
Binder's fourth-order cumulant for different values of $N$ and two
different values
of $z$. The critical noise parameter $q_{c}$, for a given value of $z$, is
estimated
as the point where the curves for different system sizes $N$ intercept
each other.
In Fig 3. the phase diagram is shown as the dependence of the critical noise
parameter $q_{c}$ on connectivity $z$ obtained from the data of Fig. 2.

The phase diagram of the majority-vote model on directed Barab\'asi-Albert
network
shows that for a given network (fixed $z$ ) the system becomes ordered for
$q<q_{c}$, whereas it has zero magnetisation for $q\geq q_{c}$. We notice
that the
increase of $q_{c}$ as a function of $z$ is slower than the one than in
\cite{brad}. In
the Fig. 4 we plot the dependence of the magnetisation at $q=q_{c}$ with
the system
size. The slopes of curves correspond to the exponent ratio $\beta/\nu$ of
according
to Eq. (5).
The results show that the exponent ratio $\beta/\nu$ decreases when $z$
increases,
see Table I.

In Fig. 5 we display the scalings for susceptibility at $\chi(q_{c}(N))$
(circle) for its maximum amplitude $\chi_{N}^{max}$, and
$\chi(q_{c}(N))$ (square) obtained from the Binder's cumulant versus $N$
for connectivity $z=8$. The exponents ratios $\gamma/\nu$ are obtained
from the slopes
of the straight lines. For almost all the values of $z$, the exponents
$\gamma/\nu$
of the two estimates disagree  (Table I). An increased $z$ means a 
tendency to
increase the exponent ratio $\gamma/\nu$, see Table I, so that they
disagree with
the results of Luiz et al \cite{brad} , where the values of exponents ratio
$\gamma/\nu$ are almost all equal and with a slight tendency to decrease.
Therefore
we cannot use the Eq.(9), for fixed $z$, obtain the critical exponent
$1/\nu$. In
the Fig. 6 we show the critical behavior of
$\beta/\nu$ and $\gamma/\nu $ as a function of connectivity $z$.

To obtain the critical exponent $1/\nu$, we calculated numerically
$U^{'}(q)=dU(q)/dq$ at the critical point for each values of $N$ at
connectivity
fixed $z$. The results
are  bad agreement with the scaling relation (7). Then, also we cannot
calculate the
exponents $1/\nu$, through this relation. Therefore we do not obtain to
get the
values of the exponents $1/\nu$ for each connectivity $z$

The Table I resumes the values of $q_{c}$, the exponents $\beta/\nu$,
$\gamma/\nu$,
and
the effective dimensionality of systems. For all values of $z$ the value
$D_{eff}=1$, which has been obtained from the Eq. (9), therefore when $z$
increases,
 $\beta/\nu$ decreases and $\gamma/\nu$ increases, thus providing the
value of
$D_{eff}=1$ (along with errors). Therefore, the directed Barab\'asi-Albert
network
has the same effective dimensionality as Erd\"os-R\'enyi's random graphs
\cite{brad}
. J. M. Oliveira \cite{mario} showed which majority-vote model  defined on
regular
lattice has critical exponents that
fall into the same class of universality as the corresponding equilibrium
Ising
model. Campos et al \cite{campos} investigated the  critical behavior of the
majority-vote on small-world networks by rewiring the two-dimensional square
lattice, Luiz et al \cite{brad} studied this model on Erd\"os-R\'enyi's
random
graphs, and Lima et al \cite{lima0} also studied this model on
Voronoi-Delaunay
lattice. The results obtained by these authors show that the critical
exponents of
majority-vote model belong to different universality classes.

\begin{table}[h]
\begin{center}
\begin{tabular}{|c c c c c c c|}
\hline
\hline
$ z $ & $q_{c}$ & $\beta/\nu$& ${\gamma/\nu}^{q_{c}}$ &
${\gamma/\nu}^{q_{c}(N)}$ &
 $ D_{eff}$\\
\hline
$ 2 $ & $ 0.434(3) $ & $ 0.477(2) $ & $ 0.064(8) $ & $ 0.895(10) $ & $
1.018(9) $ \\

$ 3 $ & $ 0.431(4) $ & $ 0.444(1) $ & $ 0.111(2) $ & $ 0.904(12) $  & $
0.999(2) $ \\
$ 4 $ & $ 0.431(3) $ & $ 0.447(1) $ & $ 0.104(2) $ & $ 0.888(9) $  & $
0.998(3) $\\
$ 6 $ & $ 0.438(2) $ & $ 0.435(2) $ & $ 0.134(5) $ & $ 0.861(3) $  & $
1.008(6) $\\
$ 8 $ & $ 0.444(5) $ & $ 0.431(1) $ & $ 0.138(2) $ & $ 0.851(5) $  & $
1.000(2) $\\
$ 10 $ & $ 0.446(3) $ & $ 0.421(2) $ & $ 0.158(3) $ & $ 0.834(7) $ & $
1.000(5) $\\
$ 20 $ & $ 0.458(4) $ & $ 0.412(1) $ & $ 0.178(2) $ & $ 0.795(11) $ &  $
1.002(2) $\\
$ 50 $ & $ 0.467(2) $ & $ 0.375(4) $ & $ 0.249(7) $ & $ 0.735(17) $ &  $
0.999(11) $\\
$ 100 $ & $ 0.474(3) $ & $ 0.363(4) $ & $ 0.269(5) $ & $ 0.674(23) $ & $
0.999(9) $\\
\hline
\hline
\end{tabular}
\end{center}
\caption{ The critical noise $q_{c}$, the critical exponents, and the
effective
dimensionality $D_{eff}$
, for  directed Barab\'asi-Albert network with connectivity $z$ .}
\label{table1}
\end{table}

Finally, we remark that our MC results obtained on directed Barab\'asi-Albert
network and undirected
(in preparation) majority-vote model show that critical exponents are
different from
the results of
\cite{mario} for regular lattice and of Luiz et al \cite{brad} for
Erd\"os-R\'enyi's
random graphs .

\bigskip

{\bf Conclusion}

In conclusion, we have presented a very simple nonequilibrium model on
directed Barab\'asi-Albert network \cite{sumour,sumourss}. Different from
the Ising model, in these networks, the Majority-vote model presents a
second-order phase transition which occurs in model with
connectivity $z>1$. The exponents obtained are differentfrom the other
models.
Nevertheless, our Monte Carlo simulations have demonstrated that the
effective dimensionality $D_{eff}$ equals units, for all values of $z$,
 that are agree with the results de Luiz et al \cite{brad}.
However, when $z$ grows, the exponents at the critical point $q_{c}$,
$\beta/\nu$
obtained by  Binder's cumulant decrease and the exponents $\gamma/\nu$  grow,
satisfying
the hyperscaling relation with $D_{eff}=1$.

  F.W.S. Lima  has the pleasure to thank D. Stauffer for many suggestions
and fruitful
discussions during the development this work and also for the revision of
this paper.
I also acknowledge the Brazilian agency FAPEPI
(Teresina-Piau\'{\i}-Brasil) for  its financial support.

\end{document}